\documentclass[twocolumn,showpacs,aps,prd,amssymb,nofootinbib,nobibnotes,floatfix,longbibliography]{aastex631}

\usepackage{newtxtext,newtxmath}
\usepackage{ae,aecompl}
\usepackage[mathscr]{euscript}

\usepackage[T1]{fontenc}
\usepackage{hyperref}
\usepackage{amsfonts}
\usepackage{amsmath}
\usepackage{mathrsfs}
\usepackage{epsfig}
\usepackage{graphicx}
\usepackage{url}
\usepackage{hyperref}
\usepackage{float}
\usepackage{color}
\usepackage[utf8]{inputenc} 
\usepackage{epsf}
\usepackage{comment}
\usepackage{slashed}
\usepackage{footmisc}
\usepackage{lipsum}
\definecolor{linkcolor}{rgb}{0.0,0.3,0.5}
\usepackage[caption=false]{subfig}

\usepackage{color}
\definecolor{cerulean}{rgb}{0.0, 0.48, 0.65}

\definecolor{navy}{rgb}{0.2, 0.0, 1.0}

\definecolor{jungle}{rgb}{0.0, 0.5, 0.0}

\usepackage{color}

\definecolor{orange}{rgb}{1,0.5,0}
\definecolor{orangeB}{rgb}{1,0.7,0}

\usepackage[mathlines]{lineno}

\begin{document}

\preprint{APS/123-QED}


\title{The Proper Motion of Strongly Lensed Binary Neutron Star Mergers in LIGO/Virgo/Kagra \\ can be Constrained by Measuring Doppler Induced Gravitational Wave Dephasing}

\author{Lorenz Zwick}
\affiliation{Niels Bohr International Academy, The Niels Bohr Institute, Blegdamsvej 17, DK-2100, Copenhagen, Denmark}

\author{Johan Samsing}
\affiliation{Niels Bohr International Academy, The Niels Bohr Institute, Blegdamsvej 17, DK-2100, Copenhagen, Denmark}

\shorttitle{Dephasing in the first detected strongly lensed NS-NS merger}
\shortauthors{Zwick et al.}

\date{\today}

\begin{abstract}
\noindent Strongly lensed binary neutron star (NS-NS) mergers are expected to be observed once LIGO/Virgo/Kagra reaches the planned A+ or proposed A\# sensitivity. We demonstrate that the relative transverse velocity of the source-lens system can be constrained by comparing the phase of the two associated gravitational wave (GW) images, using both semi-analytical and numerical Bayesian methods. For A+ sensitivity, a one-sigma NS-NS merger signal in magnification $(\mu=200)$ and redshift $(z_{\rm S}=1)$ will carry a marginally detectable dephasing signature for a source transverse velocity of $\sim 1800$ km/s. This is comparable to the velocity dispersion of large galaxy clusters. Assuming the same population distribution, the most likely source parameters of $\mu=100$ and $z_{\rm S}=1.4$ are always expected to showcase detectable dephasing imprints for A\# sensitivity, provided they are moving with transverse velocities larger than $\sim 2000$ km/s. We conclude that a first measurement of the relative transverse velocity of a source via GW dephasing methods is likely only a few years away.
\end{abstract}


\section{Introduction}\label{sec:Introduction}

\noindent The number of gravitational wave (GW) events observed by LIGO/Virgo/Kagra (LVK) is
steadily increasing \citep{2023ApJS..267...29A}, and with planned upgrades the prospect for detecting a
strongly lensed GW signal is becoming more and more
likely \citep[e.g.][]{2022ApJ...929....9X, smith2023}. Indeed, both the first
strongly lensed binary black hole (BH) and binary neutron star (NS) mergers are
expected for the beginning of the fifth LVK observation run (O5) \cite[e.g.][]{LIGOScientific:2021izm,LIGOScientific:2023bwz}. 

\begin{figure}
    \centering
    \includegraphics[width=0.45\textwidth]{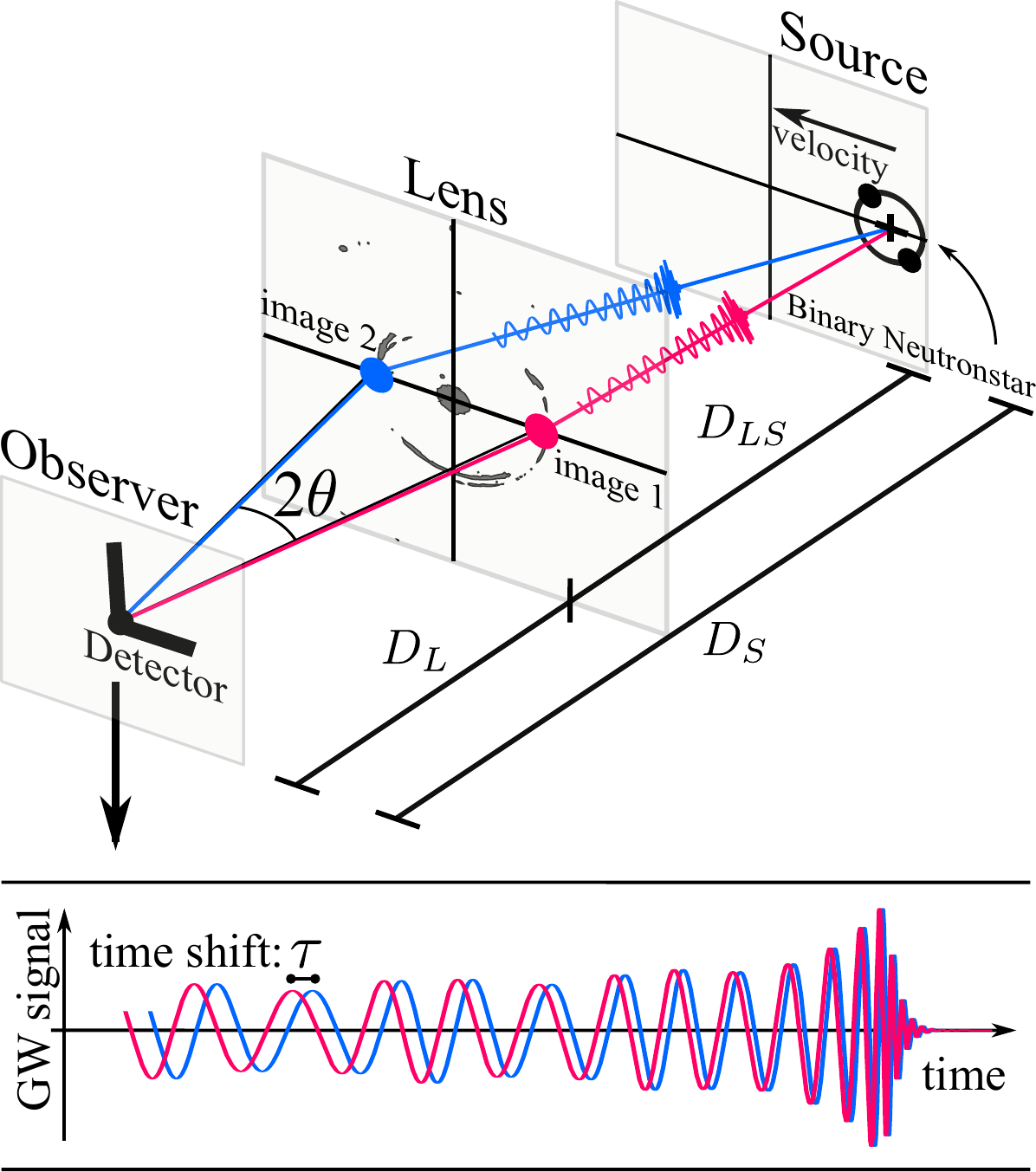}
    \caption{Sketch of the geometry of the source-lens-observer system. The different line-of-sights for the two images create a differential Doppler shift, which manifests as a phase shift between the two GW signals. In our setup, one waveform is chosen to be a reference, while the other is shifted by Eq. \ref{eq:dphi_general}.}
    \label{fig:sketch}
\end{figure}

Strong lensing of GW sources has many interesting observational prospects \citep{2003ApJ...595.1039T, Ezquiaga:2020dao, 2020arXiv200712709D, Goyal:2020bkm, Cremonese:2021puh, Tambalo:2022wlm, Jana:2022shb, 2022MNRAS.515.3299G, Caliskan:2023zqm, 2023PhRvD.108d3527T, Goyal:2023uvm, Chen:2024xal, 2024PhRvD.109b4064S, 2024PhRvD.110d4054P}.
A particularly unique one is probing the relative transverse proper
motion of the GW source by comparing the GW signals coming from the different images,
as these allow the observer to see the GW source from different
lines-of-sight (LOS) \citep[e.g.][]{2009PhRvD..80d4009I, 2024PhRvD.109b4064S, 2024arXiv241214159S}.
The different LOS give rise to a differential Doppler shift between the GW images, which manifests itself
as a relative shift in the GW phase, measurable with a sufficiently sensitive
GW detector and appropriate waveform models \citep{2019smith}.
Variations of this method have been discussed both in terms of electromagnetic signals
\citep[][]{1986A&A...166...36K, 1989Natur.341...38C, 1989LNP...330...59B, 2004PhRvD..69f3001W},
as well as for GW systems \citep[e.g.][]{2009PhRvD..80d4009I, 2020PhRvD.101h3031D,
2022MNRAS.515.3299G, 2024arXiv241016378Y, 2024PhRvD.109b4064S}, with implications 
for deci-hertz detectors, such as DECIGO/TianQin/Taiji \citep{2011CQGra..28i4011K, 2016CQGra..33c5010L, 10.1093/nsr/nwx116, 2020PhRvD.101j3027L}, and for the Einstein Telescope (ET) and
Cosmic Explorer (CE) \citep{2024arXiv241214159S} that are expected to observe $\sim$ hundreds
lensed GW sources per year \citep{2022ApJ...929....9X}.

In this letter, we show that the planned sensitivity of the advanced LVK facilities over the next years is sufficient to detect GW phase shifts from strongly lensed NS-NS GW mergers, which will allow us to directly
measure their relative transverse proper motion. In particular, we focus on the planned A+ and proposed A\# sensitivites as detailed in e.g. the \href{https://dcc.ligo.org/LIGO-T2200287/public}{T2200287v3} public document. This measurement becomes possible due
to the low mass of the NS, which results in
a relatively large number of GW cycles for phase-shift detection, combined with the possibility for
lens magnifications of order $\sim \mathcal{O}(10^2)$ \citep{smith2023}. Together with the additional
detectable GW phase shift from a possible acceleration of the NS-NS GW
source \citep[e.g.][]{Tamanini2020,tiwari2024,2024arXiv240305625S, 2024arXiv240804603H, 2024arXiv241108572H}, our study hints that { detections of GW dephasing and GW based constraints on the peculiar velocity of NS-NS merger hosts will become a reality over the next few years.}

\section{Methods}\label{sec:Methods}

\noindent A sketch of the lensed GW source and the relevant geometrical quantities is shown in Fig. \ref{fig:sketch}. In this scenario, GW dephasing between the two lensed images arises as a result of the difference in Doppler shift between the two different LOS projections. The GW phase shift for a two-image lens model is given by \citep{2009PhRvD..80d4009I, 2024arXiv241214159S},
\begin{align}
    \delta \phi \approx 4 \pi f \frac{v_{\rm Dop}}{c}\theta T_{\rm obs},
    \label{eq:dphi_gen}
\end{align}
where $f$ is the GW frequency, $2\theta$ the angular separation between the two
images, $T_{\rm obs}$ the observation time, $c$ the speed of light, $v_{\rm Dop}$ an effective
transverse velocity (projected onto the line connecting the two images) given by \citep{1986A&A...166...36K, 2009PhRvD..80d4009I, 2024arXiv241214159S},
\begin{align}
v_{\rm Dop} = v_{\rm{O}} + \frac{D_{\rm{L}}}{D_{\rm LS}}\frac{1+z_{\rm L}}{1+z_{\rm S}}v_{\rm S} - \frac{D_{\rm S}}{D_{\rm LS}}v_{\rm {L}},
\label{eq:vd}
\end{align}
$D_{i}$ are angular diameter distances (see Fig. \ref{fig:sketch}), and the $v_i$ are transverse peculiar velocities. Note that in this formula, the lens velocity is weighted roughly twice as much as the source velocity. In this work, we always take $v_{\rm O}=0$ km/s. The angular separation is given by,
\begin{align}
    \theta = \frac{4 G M_{\rm L}D_{\rm LS}}{c^2D_{\rm L}D_{\rm S}},
\end{align}
where $M_{\rm L}$ is the mass of the lens. As shown in \cite{2024arXiv241214159S}, for an equal mass chirping binary source, the
GW dephasing from Eq. \ref{eq:dphi_gen} can be written as a function of GW frequency,
\begin{align}
	\delta{\phi}   & \approx \frac{2^{4/3} 5}{\pi^{5/3} 128} \frac{c^{5}}{G^{5/3}} \times \frac{{\theta}}{m^{5/3}f^{5/3}}\frac{v_{\rm Dop}}{c}, \nonumber\\
                       & \approx 0.1\ \text{rad} \left(\frac{\theta}{25''}\right) \left(\frac{1.4\, {\rm{M}}_{\odot}}{m} \right)^{5/3} \left(\frac{10\, \rm{{Hz}}}{f}\right)^{5/3} \left(\frac{v_{\rm Dop}}{1500\ {\rm{km/s}}} \right),
    \label{eq:dphi_general}
\end{align}
where $m$ is the mass of a single NS. For these parameters, we see that the magnitude of the GW dephasing is at the threshold of what is typically considered detectable for a high SNR GW signal. Doppler velocities of more than $10^3$ km/s are expected, since the typical velocity dispersion of large galaxy clusters is of similar order \citep{2022barahona}. We take this as motivation to investigate the full phase space of parameters and determine how the presence of two images can aid in inferring the dephasing.

In our setup, the GW signal from one of the images is identified as a reference, while the GW signal from
the other image will have its phase shifted by Eq. \ref{eq:dphi_general}. We make the simplifying assumption that both reference and shifted waveforms
have comparable magnifications $\mu$, as in \cite{2024arXiv241214159S}. While this is not typically the case for strongly lensed images with magnification $\mu\sim100$, here we identify the reference magnification $\mu$ as being the geometric mean of the two separate image magnifications $\mu \sim \sqrt{\mu_1\mu_2}$. This gives an appropriate estimate for the $\delta$SNR results, unless the magnifications differ by orders of magnitudes. We leave a more thorough study of the effect of varying the two magnifications for future work, though note here that constraining them separately can aid to break degeneracies.
To model the reference GW, we employ magnified circular vacuum waveforms of the form \citep{blanchet,2007maggiore,Blanchet14}:
\begin{align}
    \label{eq:wavemod}
    \tilde{h}(f) =\sqrt{ \mu}\frac{Q}{D}\left(\frac{G\mathcal{M}_z}{c^3}\right)^{5/6}f^{-7/6}\times\sum_{lm} \mathcal{H}^{lm} \exp\left[- im\psi\right],
\end{align}
where the tilde denotes the Fourier space representation, and the mode coefficients $\mathcal{H}^{lm}$ are known to very high post-Newtonian (PN) orders \citep{Blanchet14} and the factor $Q=2/5$ accounts for the sky averaged response of an L shaped interferometer to incident GW \citep{2007maggiore,2019robson}. Other variables are listed in Eq. \ref{eq:paramssss}. The Fourier phase $\psi$ is given by:
\begin{align}
    \label{eq:wavemod2}
   \psi &= -\pi/4 + \phi_{\rm c}- 2 \pi f t  +  2\left(8\pi \mathcal{M}_zf \right)^{-5/3} \nonumber \\ &+\text{PN corrections}.
\end{align}
In this work, we use third order PN waveforms for binaries with no spin. To efficiently estimate the detectability of dephasing over a large parameter space, we employ the $\delta$SNR criterion \citep{Kocsis:2011ka}. It states that a waveform perturbation is detectable if the following inequality is satisfied:
\begin{align}
\label{eq:dSNRcrit}
   \delta \text{SNR} &\equiv \sqrt{\left<\Delta h,\Delta h \right>} > \mathcal{C}, \\
   \Delta h&= h_{\rm ref} - h_{\rm Dop},
   \label{eq:Deltah}
\end{align}
where $h_{\rm ref}$ is the waveform of the reference GW image, and $h_{\rm Dop} = h_{\rm ref} \exp(i\delta\phi)$ is the Doppler shifted waveform. The inner product $\left< \cdot , \cdot \right>$ represents the signal-to-noise (SNR):
\begin{align}
    \left< h_1 , h_2\right> = 2\int_0^{\infty} \frac{\tilde{h}_1\tilde{h}_2^{*} + \tilde{h}_1^{*}\tilde{h}_2}{S_{\rm n}(f')}{\rm{d}}f',
\end{align}
where the super script `$*$' denotes the complex conjugate, and $S_{\rm n}$ is the noise power spectral density of the detector.
Here we take the $S_{n}$ of the advanced LVK detectors, focusing on A+ \citep{2022ligopl,2025ligopl} and A\# \citep{2024ligosharp} sensitivity. Values of $\delta$SNR higher than a few indicate that a given waveform perturbation may be detectable, without accounting
for the problem of degeneracies \citep{2023zwick}. In this letter, we further employ a Bayesian parameter inference pipeline in order to investigate degeneracies and to validate the $\delta$SNR results, which is detailed in the appendix \ref{sec:MCMC}. We note that the goal of this pipeline is simply to show the advantages of jointly inferring the parameters of the reference waveform and Doppler shifted waveform. It is not intended as an exhaustive suite of tests including the effect of varying priors, waveform models, or the details of the dephasing prescriptions, which we leave to future work.

Finally, we note that the distributions of strongly lensed NS-NS mergers for the LVK O5 run, which is equivalent to A+ sensitivity, have been extensively analysed in \cite{smith2023}. Typical NS-NS mergers will be observed at $z_{\rm S} \sim 1.4$ with a magnification of $\mu \sim \mathcal{O}(10^2)$. The lensing kernel for such binaries roughly peaks at $z\sim 0.7$ and lens masses of $\sim 10^{14}$ M$_{\odot}$ \citep{2020robertson}. In this work, we will always assume that $z_{\rm L}=0.5z_{\rm S}$, though note that the dephasing becomes stronger for lenses that are closer to the source. Furthermore, the reconstructed LVK distribution of NS-NS mergers peaks at a total mass of $M\sim 3$ M$_{\odot}$ and approximately equal mass ratio \citep{LIGOScientific:2021izm,LIGOScientific:2023bwz}. Crucially, it is expected that LVK will be able to observe $\mathcal{O}({\rm{few}})$ strongly lensed NS-NS mergers over the first few years after A+ sensitivity is reached \citep{smith2023}.

\section{Results} \label{sec:Results}
\begin{figure}
    \centering
    \includegraphics[width=0.48\textwidth]{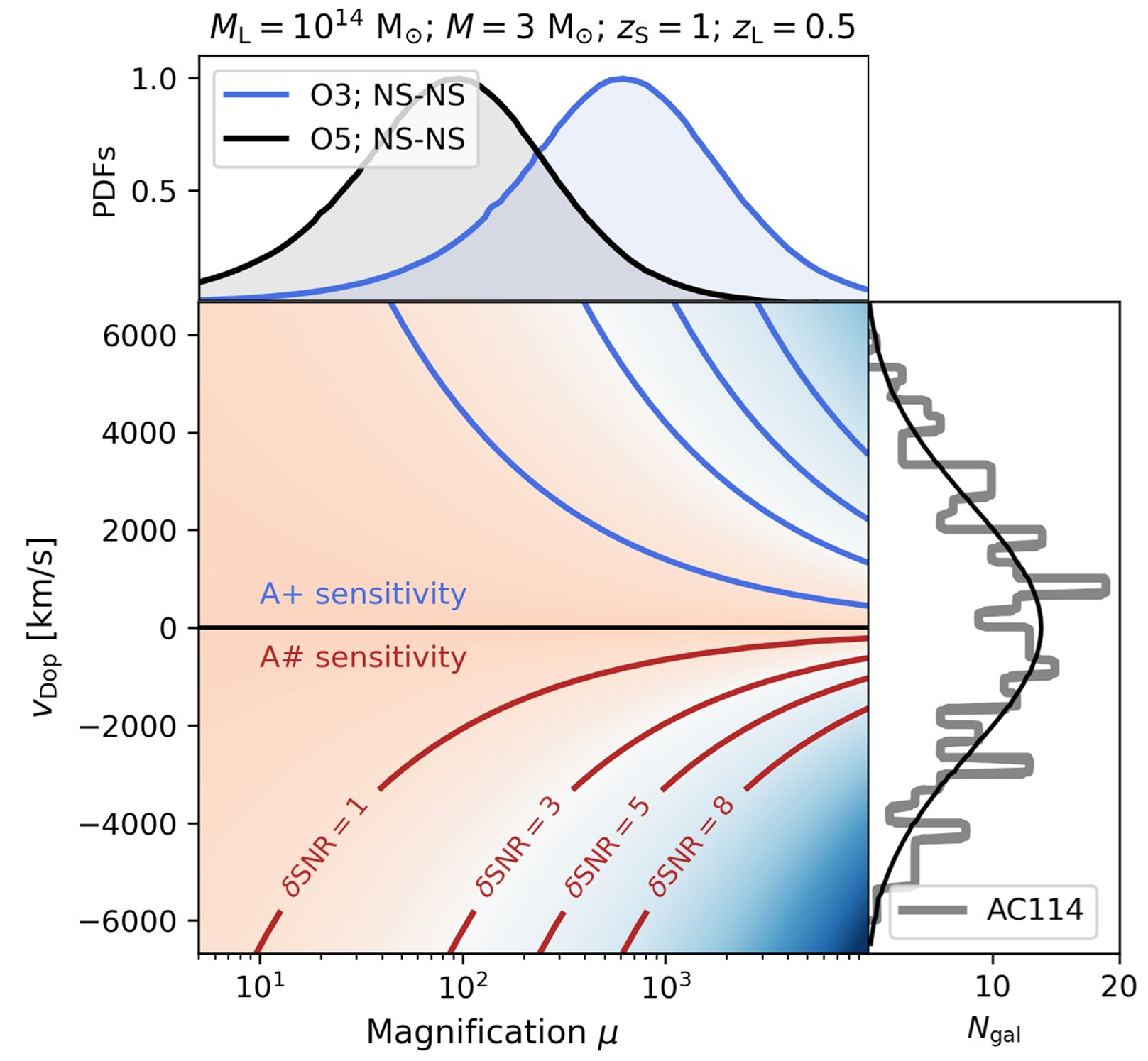}
    \caption{Contours for the $\delta$SNR of Doppler dephasing due to lensing in a strongly lensed NS-NS merger for LVK A+ and A\# sensitivity. The displayed range in relative velocity is compared to the observed line-of-sight velocity distribution of the $\sim 500$ members of the galaxy cluster AC-114 (grouped in $250$ km/s bins, and modified by the cosmological factor in Eq. \ref{eq:vd}). The magnification range is compared with theoretical predictions for {the distributions of} strongly lensed O3 and O5 sources \citep{smith2023}. {Note that the absolute amount of detected lensed sources in O5, i.e. $\mathcal{O}$(few), is  at least two order of magnitudes higher than the expected amount in O3.} All chosen parameters are typical, demonstrating how detected strongly lensed NS-NS have the potential to showcase high $\delta$SNR dephasing.}
    \label{fig:dSNR}
\end{figure}
\noindent Fig. \ref{fig:dSNR} shows the contours for the $\delta$SNR of Doppler dephasing as a function of the magnification and the relative Doppler velocity. The range of magnifications is compared to the expected distributions for the O3 and O5 LVK observation runs. The most likely magnification for O5 is $\mu \sim 100$, with an approximately a log-normal distribution with standard deviation of $\sim 0.3$ dex. For a one-sigma source in magnification ($\mu \sim 200$) at $z_{\rm S}=1$ and A+ sensitivity, the dephasing reaches  a $\delta$SNR of $\sim 1$ at $v_{\rm Dop} \sim 2600$ km/s, corresponding to a source transverse velocity of $v_{\rm S} \sim 1800$ km/s as in Eq. \ref{eq:Deltah}. As a comparison, we show a histogram of the observed velocities of $\sim 500$ galaxies in AC-114, a galaxy cluster with a velocity dispersion of $\sigma = 1893$ km/s \citep{2004ac114,2015ac114}. As we can see, a combination of magnification and Doppler velocities that produces large $\delta$SNR, i.e$>3$, is possible for a significant region in parameter space, especially for A\# sensitivity.

\begin{figure}
    \centering
    \includegraphics[width=0.47\textwidth]{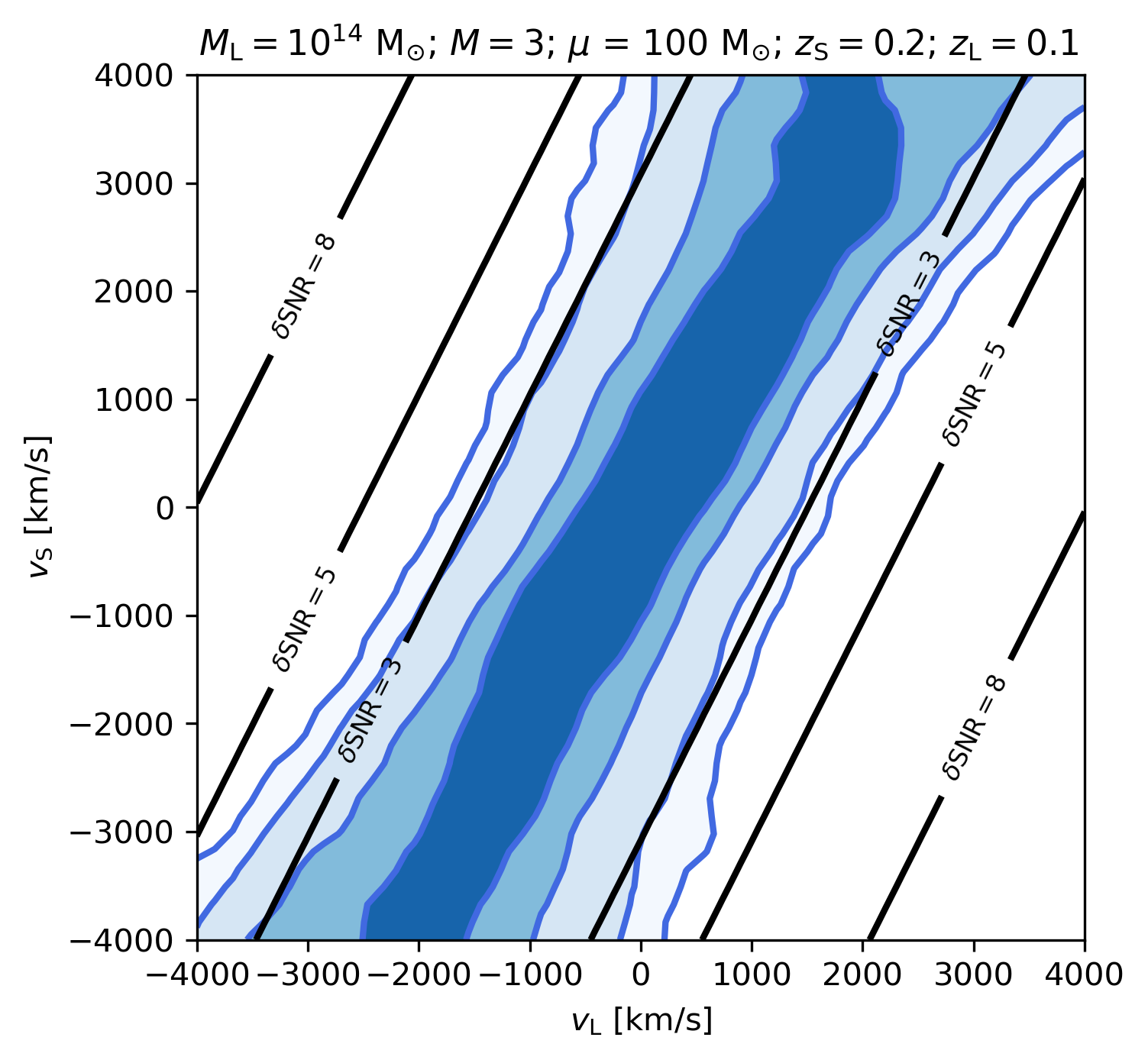}
    \caption{Contours of the $\delta$SNR for the Doppler induced dephasing of a strongly lensed NS-NS mergers for A+ sensitivity. The countours are compared with the marginalised posterior (blue) from a joint parameter inference test over [$z_{\rm S}$,$\mathcal{M}$,$\nu$,$\phi_c$,$\mu$,$z_{\rm L}$], where the lens mass is assumed to be $M_{\rm L}= 10^{14}$ M$_{\odot}$, proving the robustness of the $\delta$SNR criterion. The posterior contour levels show (0.01,0.1,0.3,0.6) of the maximum posterior likelihood. The one-sigma 2-d contour containing $\sim$ 50\% of the samples roughly corresponds to a $\delta$SNR of 1. This particular MCMC test is representative of many other with various injected parameters.}
    \label{fig:MCMC}
\end{figure}

In Fig. \ref{fig:MCMC}, we contrast the $\delta$SNR contours as a function of the source and lens velocities with the marginalized posterior distribution of a parameter inference test for a high SNR source in which we injected zero dephasing. In this particular test, we fixed the lens mass to be $M_{\rm L}=10^{14}$ M$_{\odot}$ to break the $\theta\times v_{\rm Dop}$ degeneracy and marginalised over the remaining parameters. We observe that the parameter inference precisely follows the expectations set by the $\delta$SNR criterion, being able to exclude large values of the relative velocity and recovering the degeneracy between source and lens velocities. In particular, the posterior distribution strongly disfavours regions of parameter space that roughly correspond to the limits $\delta$SNR$\sim$few. In particular, the contour containing 50 percent of the samples (i.e. the 2-d, 1 sigma contour) roughly corresponds to the criterion $\delta$SNR = 1. This is a highly non-trivial result, considering that the posteriors are marginalised over a total of six ulterior parameters, i.e. [$z_{\rm S}$,$\mathcal{M}$,$\nu$,$\phi_c$,$\mu$,$z_{\rm L}$]. It arises as a  consequence of the unique advantage of having two separate images that have already been identified as signals, and the parameters of which can be inferred jointly \citep[see e.g.][for strategies to identify lensed signals]{2022more,2023jose}.
We note here that several degeneracies and correlations exist between the parameters of the dephasing prescription, and only a combination of $\theta \times v_{\rm Dop}$ can be realistically extracted from a pure GW signal. Nevertheless, once the lens is identified with a known galaxy or galaxy cluster, the lens mass will be known and so will the angle $\theta$. The latter can also be reconstructed by fitting a combination of the delay time and two different magnifications for each GW image, though this requires a more sophisticated lens model \citep{2025arXiv250102096V}. Furthermore, the lensing kernel strongly peaks around $10^{14}$ M$_{\odot}$, which can be implemented as a prior. Then, the Doppler velocity $v_{\rm D}$ and perhaps the individual components $v_{\rm L}$ and $v_{\rm S}$ can be separated out. Further opportunities to break degeneracies can be afforded by employing multi-messenger follow-up observations. All of these aspects will not be discussed here, though the many opportunities deserve to be studied in detail in further work. Crucially, they are likely to further improve the possible constraints on the source-lens relative motion, meaning that our estimates are in all likelihood conservative.

\begin{figure}
    \centering
    \includegraphics[width=0.45
    \textwidth]{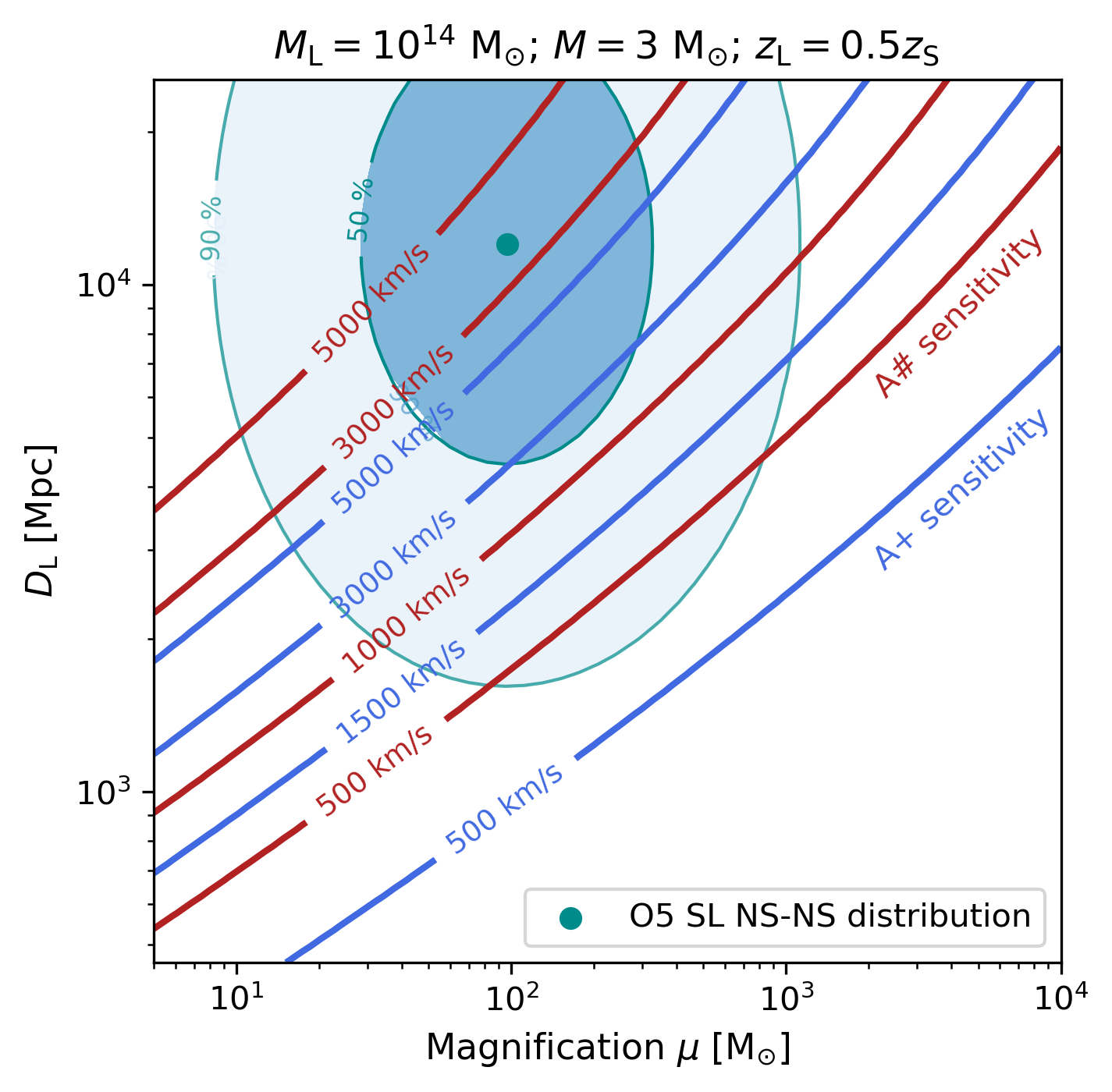}
    \caption{We show the 50\% and the 90\% contours of the expected distributions of strongly lensed NS-NS mergers (cyan) for A+ sensitivity estimated from \citep{smith2023} and assuming log-normal distributions. The velocity contours denote the critical transverse source velocity $v_{\rm crit}$ to achieve marginally detectable dephasing (see text) for A+ (blue) and A\# (red) sensitivity. In other words, they denote the possible constraint on the transverse source velocity via a GW dephasing measurement.}
    \label{fig:poster}
\end{figure}

In Fig. \ref{fig:poster} we compare the 50\% and the 90\% contours of the expected distributions of strongly lensed NS-NS mergers in luminosity distance and magnification for A+ sensitivity \citep{smith2023} with the critical source velocity $v_{\rm crit}$. The latter is defined as the absolute value of the source transverse velocity required to achieve a value of $\delta$SNR=1, which is sufficient for a marginal detection according to our Bayesian analysis (see Fig. \ref{fig:MCMC}). Roughly 15\% (5\%) of expected sources will have a critical velocity lower than  3000 km/s (1500 km/s) for A+ sensitivity.
Typical galaxy clusters have velocity dispersions of $\sigma \sim 1000$ km/s \citep{2022barahona}, though often exceed it as in the case of e.g. AC-114 ($\sigma=1893$ km/s). Thus, the velocity dispersions of large galaxy clusters are comparable to the expected critical velocity $v_{\rm rit}$ for a small but substantial fraction of strongly lensed NS-NS mergers for A+ sensitivity. While the expected distributions for A\# have not been studied yet, the critical velocity contours are extremely promising, indicating that up to half of the expected signals may have detectable dephasing imprints. Therefore, it is likely that out of the first few detected strongly lensed NS-NS signals in LVK, at least one can can be used to measure the relative velocity of its host galaxy via a measurement of GW dephasing between the two lensed images. Non-detections of dephasing can instead be used constraints on the relative velocity of the lens-source system.

\section{Conclusions}\label{sec:Conclusions}
\noindent
Strongly lensed NS-NS mergers are expected to be detected in the first years after the LVK facilities reach A+ sensitivity. We have shown that, for a significant fraction of the expected source parameter phase space, such signals will showcase detectable GW dephasing signatures due to the peculiar transverse velocity of the lens and source. The prospects are even more promising when considering the proposed A\# sensitivity curve. {The methodology proposed here allows for a "dark", or GW based, constraint of the host galaxy transverse velocity within its galaxy cluster. Additionally, not modelling the effects here presented can lead to the misidentification of a lensed GW signal as two distinct sources with different chirp mass, due to the $f^{-5/3}$ scaling of the Doppler induced dephasing.} We conclude that the first detection of GW dephasing and the first use of GW measurements to directly infer information of the source's environment is likely only a few years away. {In light of this, We suggest that further investigating the synergies between GW dephasing measurements and multi-messenger observations in lensed NS-NS mergers should be prioritised.}

\section{Acknowledgments}
L.Z. and J.S. are supported by the Villum Fonden grant No. 29466, and by the ERC Starting Grant no. 101043143 -- BlackHoleMergs led by J. Samsing. The authors are grateful Juan Urrutia, Mikołaj Korzyński, Miguel Zumalacárregui, Jose María Ezquiaga, Rico K. L. Lo, Luka Vujeva, and Graham Smith, for useful discussions. J.S. further thanks The Erwin Schrödinger International Institute
for Mathematics and Physics (ESI) and the organizers of the workshop "Lensing and Wave Optics in Strong Gravity"
where part of this work was carried out.
L.Z. acknowledges A.O. for at vaere et hyggemonster. L.Z acknowledges FanaticoClimb for being a great guesthouse to write papers.

\appendix
\section{MCMC pipeline}
\label{sec:MCMC}
\begin{figure}
    \centering
    \includegraphics[width=0.9
    \textwidth]{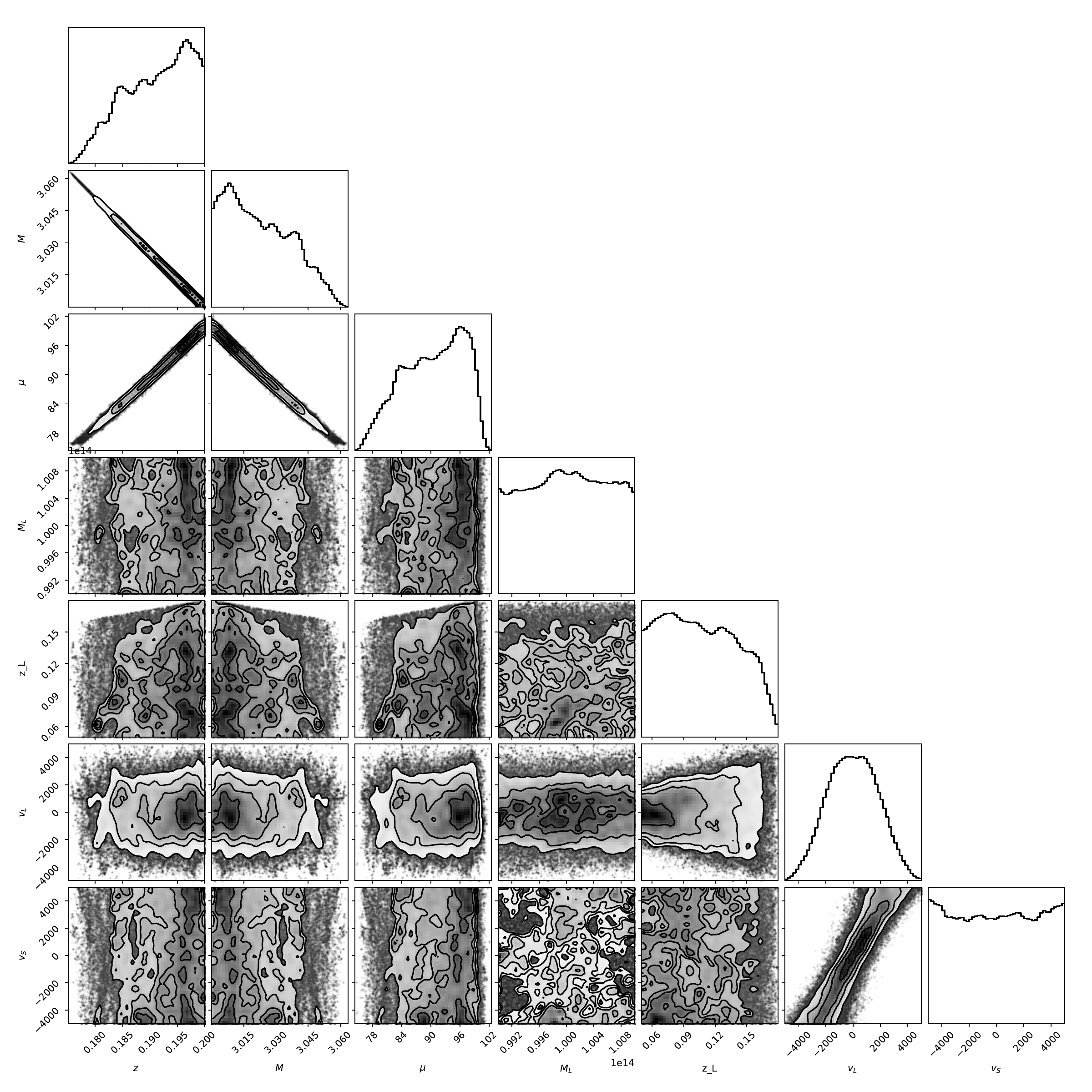}
    \caption{Full corner plot of a numerical parameter inference run. The injected parameters are $z_{\rm S}=0.2$, $z_{\rm L}=0.1$, $M_{\rm L}=10^{14}$ M$_{\odot}$, $M=3$, $q=1$ and zero Doppler velocity. The results are representative of other numerical tests with different injected parameters.}
    \label{fig:fullMCMC}
\end{figure}

\noindent For our numerical paramter inference tests, we limit our analysis to the binary's intrinsic parameters and initial phase, while performing a sky average of a source's position on the sky and orientation. Furthermore, we assume that the binary coalescence time may be measured with high precision through the merger and ring-down portions of the GW signal, which we do not model here. Knowing the merger time of both images allows to infer and account for the time delay caused by the lens, which in turn allows to precisely define the reference and Doppler shifted waveforms. Therefore, the parameters that we wish to infer are:
\begin{align}
\label{eq:paramssss}
    \text{Ref. par.:} = &\begin{cases}
        z_{\rm S}; & \text{ Source redshift}\\
        \mathcal{M}; & \text{Chirp mass}\\
        \nu; & \text{Symmetric mass ratio} \\
        \phi_{\rm c}; & \text{Initial binary phase} \\
        \mu; & \text{Magnification}
    \end{cases}\\ \nonumber \\
     \text{Deph. par.:} = &\begin{cases}
        z_{\rm L}; & \text{Lens redshift}\\
        M_{\rm L}; & \text{Lens mass}\\
        v_{\rm L}; & \text{Lens velocity}
        \\
        v_{\rm S}; & \text{Source velocity}
    \end{cases}
\end{align}
To perform the parameter estimation we use Monte-Carlo-Markhov-Chain (MCMC) methods.
The goal of MCMC based parameter estimation is to sample the posterior distributions of a set of waveform parameters $\Theta$ given a likelihood function $\mathcal{L}$ of the form:
\begin{align}
    \mathcal{L}\left(h;\Theta \right) \propto \exp\left[ - \big(h(\Theta) - h(\Theta_{\rm GT}),h(\Theta) - h(\Theta_{\rm GT}) \big) \right],
\end{align}
where $h$ is the particular waveform model used for both the injected signal and the trials. Here $\Theta_{\rm GT}$ is the vector of "ground truth" parameters chosen for the injected signal. The detector noise is treated as being stationary, representing the average value for the noise power spectral density over many realizations. Given that we have at least two images, we are able to simultaneously infer the parameters of both the reference and Doppler shifted waveforms. The appropriate likelihood is therefore given by:
\begin{align}
    \mathcal{L}_{\rm{joint}} \propto \mathcal{L}(h_{\rm ref};\Theta)\times\mathcal{L}(h_{\rm Dop};\Theta).
\end{align}
We use the affine invariant sampler \texttt{emcee} \citep{2013emcee} to perform the numerical tests, typically running 27 parallel walkers for approximately 30'000 to 50'000 steps, where the typical auto-correlation time of the walkers is $\sim 200$. We initialize the walkers in the vicinity of the injected true values in order to speed up convergence and to avoid having to specify priors for the vacuum binary parameters (which is appropriate only for high SNR sources). The priors for the perturbation parameters are flat within appropriate ranges and do not influence our results unless stated otherwise. An example of a full paramter inference run is shown in Fig \ref{fig:fullMCMC}.

\bibliographystyle{aasjournal}
\bibliography{main}

\begin{thebibliography}{}
\expandafter\ifx\csname natexlab\endcsname\relax\def\natexlab#1{#1}\fi
\providecommand{\url}[1]{\href{#1}{#1}}
\providecommand{\dodoi}[1]{doi:~\href{http://doi.org/#1}{\nolinkurl{#1}}}
\providecommand{\doeprint}[1]{\href{http://ascl.net/#1}{\nolinkurl{http://ascl.net/#1}}}
\providecommand{\doarXiv}[1]{\href{https://arxiv.org/abs/#1}{\nolinkurl{https://arxiv.org/abs/#1}}}

\bibitem[{Abbott {et~al.}(2021)}]{LIGOScientific:2021izm}
Abbott, R., {et~al.} 2021, Astrophys. J., 923, 14, \dodoi{10.3847/1538-4357/ac23db}

\bibitem[{{Abbott} {et~al.}(2023){Abbott}, {Abe}, {Acernese}, {Ackley}, {Adhicary}, {Adhikari}, {Adhikari}, {Adkins}, {Adya}, {Affeldt}, {Agarwal}, {Agathos}, {Aguiar}, {Aiello}, {Ain}, {Ajith}, {Akutsu}, {Albanesi}, {Alfaidi}, {Al-Jodah}, {All{\'e}n{\'e}}, {Allocca}, {Almualla}, {Altin}, {Amato}, {Amez-Droz}, {Amorosi}, {Anand}, {Ananyeva}, {Andersen}, {Anderson}, {Anderson}, {Andia}, {Ando}, {Andrade}, {Andres}, {Andr{\'e}s-Carcasona}, {Andri{\'c}}, {Ansoldi}, {Antelis}, {Antier}, {Aoumi}, {Apostolatos}, {Appavuravther}, {Appert}, {Apple}, {Arai}, {Araya}, {Araya}, {Areeda}, {Ar{\`e}ne}, {Aritomi}, {Arnaud}, {Arogeti}, {Aronson}, {Arun}, {Asada}, {Ashton}, {Aso}, {Assiduo}, {Assis de Souza Melo}, {Aston}, {Astone}, {Aubin}, {Aultoneal}, {Babak}, {Badalyan}, {Badaracco}, {Badger}, {Bae}, {Bagnasco}, {Bai}, {Baier}, {Baiotti}, {Baird}, {Bajpai}, {Baka}, {Ball}, {Ballardin}, {Ballmer}, {Baltus}, {Banagiri}, {Banerjee}, {Bankar}, {Baral}, {Barayoga}, {Barber}, {Barish}, {Barker}, {Barneo}, {Barone}, {Barr},
  {Barsotti}, {Barsuglia}, {Barta}, {Barthelmy}, {Barton}, {Bartos}, {Basak}, {Basalaev}, {Bassiri}, {Basti}, {Bawaj}, {Bayley}, {Baylor}, {Bazzan}, {B{\'e}csy}, {Bedakihale}, {Beirnaert}, {Bejger}, {Bell}, {Benedetto}, {Beniwal}, {Benoit}, {Bentley}, {Yaala}, {Bera}, {Berbel}, {Bergamin}, {Berger}, {Bernuzzi}, {Beroiz}, {Berry}, {Bersanetti}, {Bertolini}, {Betzwieser}, {Beveridge}, {Bevins}, {Bhandare}, {Bhandari}, {Bhardwaj}, {Bhatt}, {Bhattacharjee}, {Bhaumik}, {Bianchi}, {Bilenko}, {Bilicki}, {Billingsley}, {Bini}, {Birnholtz}, {Biscans}, {Bischi}, {Biscoveanu}, {Bisht}, {Biswas}, {Bitossi}, {Bizouard}, {Blackburn}, {Blair}, {Blair}, {Blair}, {Bobba}, {Bode}, {Bo{\"e}r}, {Bogaert}, {Boileau}, {Boldrini}, {Bolingbroke}, {Bonavena}, {Bondarescu}, {Bondu}, {Bonilla}, {Bonilla}, {Bonnand}, {Booker}, {Bork}, {Boschi}, {Bose}, {Bose}, {Bossilkov}, {Boudart}, {Bouffanais}, {Bozzi}, {Bradaschia}, {Brady}, {Braglia}, {Branch}, {Branchesi}, {Brau}, {Breschi}, {Briant}, {Brillet}, {Brinkmann}, {Brockill}, {Brooks},
  {Brooks}, {Brown}, {Brunett}, {Bruno}, {Bruntz}, {Bryant}, {Bucci}, {Buchanan}, {Bulashenko}, {Bulik}, {Bulten}, {Buonanno}, {Burtnyk}, {Buscicchio}, {Buskulic}, {Buy}, {Byer}, {Cabourn Davies}, {Cabras}, {Cabrita}, {Cadonati}, {Caesar}, {Cagnoli}, {Cahillane}, {Calder{\'o}n Bustillo}, {Callaghan}, {Callister}, {Calloni}, {Camp}, {Canepa}, {Caneva Santoro}, {Cannavacciuolo}, {Cannon}, {Cao}, {Cao}, {Capistran}, {Capocasa}, {Capote}, {Carapella}, {Carbognani}, {Carlassara}, {Carlin}, {Carpinelli}, {Carter}, {Carullo}, {Casanueva Diaz}, {Casentini}, {Castaldi}, {Castro-Lucas}, {Caudill}, {Cavagli{\`a}}, {Cavalieri}, {Cella}, {Cerd{\'a}-Dur{\'a}n}, {Cesarini}, {Chaibi}, {Chakalis}, {Chalathadka Subrahmanya}, {Champion}, {Chan}, {Chan}, {Chandra}, {Chang}, {Chang}, {Chanial}, {Chao}, {Chapman-Bird}, {Charlton}, {Charlton}, {Chassande-Mottin}, {Chastain}, {Chatterjee}, {Chatterjee}, {Chatterjee}, {Chaturvedi}, {Chaty}, {Chatziioannou}, {Chen}, {Chen}, {Chen}, {Chen}, {Chen}, {Chen}, {Chen}, {Chen}, {Cheng},
  {Chessa}, {Cheung}, {Chia}, {Chiadini}, {Chiang}, {Chiang}, {Chiarini}, {Chiba}, {Chiba}, {Chierici}, {Chincarini}, {Chiofalo}, {Chiummo}, {Choudhary}, {Christensen}, {Chua}, {Chung}, {Ciani}, {Ciecielag}, {Cie{\'s}lar}, {Cifaldi}, {Ciobanu}, {Ciolfi}, {Clara}, {Clark}, {Clarke}, {Clearwater}, {Clesse}, {Cleva}, {Coccia}, {Codazzo}, {Cohadon}, {Colleoni}, {Collette}, {Colombo}, {Colpi}, {Compton}, {Conti}, {Cooper}, {Corban}, {Corbitt}, {Cordero-Carri{\'o}n}, {Corezzi}, {Cornish}, {Corsi}, {Cortese}, {Coschizza}, {Cottingham}, {Coughlin}, {Coulon}, {Countryman}, {Coupechoux}, {Cousins}, {Couvares}, {Coward}, {Cowart}, {Cowburn}, {Coyne}, {Coyne}, {Craig}, {Creighton}, {Creighton}, {Criswell}, {Crockett-Gray}, {Croquette}, {Crowder}, {Cudell}, {Cullen}, {Cumming}, {Cummings}, {Cuoco}, {Cury{\l}o}, {Dabadie}, {Dal Canton}, {Dall'Osso}, {D{\'a}lya}, {D'Angelo}, {Danilishin}, {D'Antonio}, {Danzmann}, {Darroch}, {Darsow-Fromm}, {Dasgupta}, {Datrier}, {Datta}, {Dattilo}, {Dave}, {Davenport}, {Davier}, {Davis},
  {Davis}, {Daw}, {Dax}, {Debra}, {Deenadayalan}, {Degallaix}, {de Laurentis}, {Del{\'e}glise}, {Del Favero}, {de Lillo}, {de Lillo}, {Dell'Aquila}, {Del Pozzo}, {de Matteis}, {D'Emilio}, {Demos}, {Dent}, {Depasse}, {de Pietri}, {De Rosa}, {de Rossi}, {Desalvo}, {de Simone}, {Dhurandhar}, {Diab}, {Diamond}, {D{\'\i}az}, {Didio}, {Dietrich}, {di Fiore}, {di Fronzo}, {di Giorgio}, {di Giovanni}, {di Giovanni}, {di Girolamo}, {Diksha}, {di Lieto}, {di Michele}, {di Pace}, {di Palma}, {di Renzo}, {Divyajyoti}, {Dmitriev}, {Doctor}, {Dohmen}, {Doleva}, {Donahue}, {D'Onofrio}, {Donovan}, {Dooley}, {Dooney}, {Doravari}, {Dorosh}, {Drago}, {Driggers}, {Drori}, {Ducoin}, {Dunn}, {Dupletsa}, {Durante}, {D'Urso}, {Duverne}, {Dwyer}, {Eassa}, {Easter}, {Ebersold}, {Eckhardt}, {Eddolls}, {Edelman}, {Edo}, {Edy}, {Effler}, {Eichholz}, {Eisenmann}, {Eisenstein}, {Ejlli}, {Engelby}, {Engl}, {Errico}, {Essick}, {Estell{\'e}s}, {Estevez}, {Etzel}, {Evans}, {Evans}, {Evans}, {Evstafyeva}, {Ewing}, {Fabrizi}, {Faedi}, {Fafone},
  {Fair}, {Fairhurst}, {Fan}, {Fan}, {Farah}, {Farr}, {Farr}, {Fauchon-Jones}, {Favaro}, {Favata}, {Fays}, {Feicht}, {Fejer}, {Fenyvesi}, {Ferguson}, {Fernandez-Galiana}, {Ferrante}, {Ferreira}, {Fidecaro}, {Figura}, {Fiori}, {Fiori}, {Fishbach}, {Fisher}, {Fittipaldi}, {Fiumara}, {Flaminio}, {Fleischer}, {Fleming}, {Floden}, {Fong}, {Font}, {Fornal}, {Forsyth}, {Franke}, {Frasca}, {Frasconi}, {Freed}, {Frei}, {Freise}, {Freitas}, {Frey}, {Fritschel}, {Frolov}, {Fronz{\'e}}, {Fujimoto}, {Fukunaga}, {Fulda}, {Fyffe}, {Gabbard}, {Gabella}, {Gadre}, {Gaglani}, {Gair}, {Gais}, {Galaudage}, {Gallardo}, {Gamba}, {Ganapathy}, {Ganguly}, {Gao}, {Gaonkar}, {Garaventa}, {Garcia-Bellido}, {Garc{\'\i}a-N{\'u}{\~n}ez}, {Garc{\'\i}a-Quir{\'o}s}, {Gardner}, {Gargiulo}, {Garufi}, {Gasbarra}, {Gateley}, {Gayathri}, {Gemme}, {Gennai}, {George}, {Gerberding}, {Gergely}, {Ghonge}, {Ghosh}, {Ghosh}, {Ghosh}, {Ghosh}, {Ghosh}, {Giacoppo}, {Giaime}, {Giardina}, {Gibson}, {Gier}, {Giri}, {Gissi}, {Gkaitatzis}, {Glanzer}, {Gleckl},
  {Glotin}, {Godfrey}, {Godwin}, {Goetz}, {Goetz}, {Golomb}, {Goncharov}, {Gonz{\'a}lez}, {Gosselin}, {Gouaty}, {Gould}, {Goyal}, {Grace}, {Grado}, {Graham}, {Granata}, {Granata}, {Gras}, {Grassia}, {Gray}, {Gray}, {Greco}, {Green}, {Green}, {Green}, {Green}, {Gretarsson}, {Gretarsson}, {Griffith}, {Griffiths}, {Griggs}, {Grignani}, {Grimaldi}, {Grote}, {Gruson}, {Guerra}, {Guetta}, {Guidi}, {Guimaraes}, {Gulati}, {Gulminelli}, {Gunny}, {Guo}, {Guo}, {Gupta}, {Gupta}, {Gupta}, {Gupta}, {Gupta}, {Gupta}, {Gurs}, {Gushima}, {Gustafson}, {Gutierrez}, {Guzman}, {Haegel}, {Hain}, {Haino}, {Halim}, {Hall}, {Hamilton}, {Hammond}, {Han}, {Haney}, {Hanks}, {Hanna}, {Hannam}, {Hannuksela}, {Hansen}, {Hanson}, {Harada}, {Harder}, {Haris}, {Harmark}, {Harms}, {Harry}, {Harry}, {Hartwig}, {Haskell}, {Haster}, {Hathaway}, {Haughian}, {Hayakawa}, {Hayama}, {Hayes}, {Healy}, {Heffernan}, {Heidmann}, {Heintze}, {Heinze}, {Heinzel}, {Heitmann}, {Hellman}, {Hello}, {Helmling-Cornell}, {Hemming}, {Hendry}, {Heng}, {Hennes},
  {Hennig}, {Hennig}, {Henshaw}, {Hernandez Vivanco}, {Heurs}, {Hewitt}, {Higginbotham}, {Hild}, {Hill}, {Himemoto}, {Hines}, {Hirata}, {Hirose}, {Ho}, {Hochheim}, {Hofman}, {Hohmann}, {Holcomb}, {Holland}, {Holley-Bockelmann}, {Hollows}, {Holmes}, {Holt}, {Holz}, {Hong}, {Hornung}, {Hoshino}, {Hough}, {Hourihane}, {Howell}, {Howell}, {Hoy}, {Hoyland}, {Hsieh}, {Hsieh}, {Hsiung}, {Hsu}, {Hu}, {Hu}, {Huang}, {Huang}, {Huang}, {Huang}, {H{\"u}bner}, {Huddart}, {Hughey}, {Hui}, {Hui}, {Husa}, {Huttner}, {Huxford}, {Huynh-Dinh}, {Hyland}, {Iakovlev}, {Iandolo}, {Idzkowski}, {Iess}, {Inayoshi}, {Inoue}, {Iorio}, {Iosif}, {Irwin}, {Isi}, {Ismail}, {Itoh}, {Iyer}, {Jaberianhamedan}, {Jacqmin}, {Jacquet}, {Jadhav}, {Jadhav}, {Jain}, {Jain}, {James}, {Jan}, {Jani}, {Janiurek}, {Janquart}, {Janssens}, {Janthalur}, {Jaraba}, {Jaranowski}, {Jarov}, {Jasal}, {Jaume}, {Javed}, {Jenkins}, {Jenner}, {Jennings}, {Jia}, {Jiang}, {Liu}, {Jin}, {Johansmeyer}, {Johns}, {Johnson}, {Johnston}, {Johny}, {Jones}, {Jones}, {Jones},
  {Jones}, {Jones}, {Joshi}, {Ju}, {Jung}, {Junker}, {Juste}, {Kajita}, {Kalaghatgi}, {Kalogera}, {Kamai}, {Kamiizumi}, {Kanda}, {Kandhasamy}, {Kang}, {Kanner}, {Kapadia}, {Kapasi}, {Karat}, {Karathanasis}, {Karki}, {Kasamatsu}, {Kas-Danouche}, {Kashyap}, {Kasprzack}, {Kastaun}, {Kato}, {Katsanevas}, {Katsavounidis}, {Katsuren}, {Katzman}, {Kaur}, {Kawabe}, {Kawazoe}, {K{\'e}f{\'e}lian}, {Keitel}, {Kellard}, {Kelley-Derzon}, {Kennington}, {Key}, {Khadka}, {Khalili}, {Khan}, {Khanam}, {Khazanov}, {Khursheed}, {Kijbunchoo}, {Kim}, {Kim}, {Kim}, {Kim}, {Kim}, {Kim}, {Kim}, {Kim}, {Kimball}, {Kimura}, {Kinley-Hanlon}, {Kirchhoff}, {Kissel}, {Kiyota}, {Klimenko}, {Klinger}, {Knee}, {Knust}, {Kobayashi}, {Koch}, {Koehlenbeck}, {Koekoek}, {Kohri}, {Kokeyama}, {Koley}, {Koliadko}, {Kolitsidou}, {Kolstein}, {Kondrashov}, {Kong}, {Kontos}, {Korobko}, {Kossak}, {Kouvatsos}, {Kovalam}, {Koyama}, {Kozak}, {Kranzhoff}, {Kranzhoff}, {Kringel}, {Krishnendu}, {Kr{\'o}lak}, {Kuehn}, {Kuijer}, {Kukihara}, {Kulkarni}, {Kumar},
  {Kumar}, {Kumar}, {Kumar}, {Kumar}, {Kume}, {Kuns}, {Kuroyanagi}, {Kuwahara}, {Kwak}, {Lacaille}, {Lagabbe}, {Laghi}, {Lakkis}, {Lalande}, {Lalleman}, {Lamberts}, {Landry}, {Lane}, {Lang}, {Lange}, {Lantz}, {La Rana}, {La Rosa}, {Lartaux-Vollard}, {Lasky}, {Lawrence}, {Laxen}, {Lazzarini}, {Lazzaro}, {Leaci}, {Leavey}, {Lebohec}, {Lecoeuche}, {Lee}, {Lee}, {Lee}, {Lee}, {Lee}, {Lee}, {Lee}, {Legred}, {Lehmann}, {Lehner}, {Lema{\^\i}tre}, {Lenti}, {Leonardi}, {Leonova}, {Leroy}, {Letendre}, {Lethuillier}, {Levesque}, {Levin}, {Leyde}, {Li}, {Li}, {Li}, {Li}, {Lin}, {Lin}, {Lin}, {Lin}, {Lin}, {Lin}, {Lin}, {Lin}, {Linde}, {Linker}, {Littenberg}, {Liu}, {Liu}, {Llamas}, {Lo}, {Lo}, {London}, {Longo}, {Lopez}, {Lopez Portilla}, {Lorenzini}, {Loriette}, {Lormand}, {Losurdo}, {Lott}, {Lough}, {Loughlin}, {Lousto}, {Lovelace}, {Lowry}, {L{\"u}ck}, {Lumaca}, {Lundgren}, {Lung}, {Lussier}, {Lynam}, {Ma}, {Ma}, {Ma'Arif}, {Macas}, {Macinnis}, {MacLeod}, {MacMillan}, {Macquet}, {Maga{\~n}a Hernandez}, {Magazz{\`u}},
  {Magee}, {Maggiore}, {Magnozzi}, {Mahesh}, {Mahesh}, {Maini}, {Majorana}, {Makarem}, {Maliakal}, {Malik}, {Man}, {Mandic}, {Mangano}, {Mannix}, {Mansell}, {Mansingh}, {Manske}, {Mantovani}, {Mapelli}, {Marchesoni}, {Pina}, {Marion}, {M{\'a}rka}, {M{\'a}rka}, {Markakis}, {Markosyan}, {Markowitz}, {Maros}, {Marquina}, {Marsat}, {Martelli}, {Martin}, {Martin}, {Martinez}, {Martinez}, {Martinez}, {Martinez}, {Martinovic}, {Martynov}, {Marx}, {Masalehdan}, {Mason}, {Masserot}, {Reid}, {Mastrodicasa}, {Mastrogiovanni}, {Mateu-Lucena}, {Matiushechkina}, {Matsunaga}, {Mavalvala}, {McCarthy}, {McClelland}, {McClincy}, {McCormick}, {McCuller}, {McGhee}, {McGinn}, {McIsaac}, {McIver}, {McLeod}, {McRae}, {McWilliams}, {Meacher}, {Mehmet}, {Mehta}, {Meijer}, {Melatos}, {Mendell}, {Menendez-Vazquez}, {Menoni}, {Mercer}, {Mereni}, {Merfeld}, {Merilh}, {Merritt}, {Merzougui}, {Messenger}, {Messick}, {Meyers}, {Meylahn}, {Mhaske}, {Miani}, {Miao}, {Michaloliakos}, {Michel}, {Michimura}, {Middleton}, {Mihaylov}, {Miller},
  {Miller}, {Miller}, {Miller}, {Millhouse}, {Mills}, {Milotti}, {Minenkov}, {Mio}, {Mir}, {Miravet-Ten{\'e}s}, {Mishra}, {Mishra}, {Mishra}, {Mistry}, {Mitchell}, {Mitra}, {Mitrofanov}, {Mitselmakher}, {Mittleman}, {Miyakawa}, {Miyoki}, {Mo}, {Modafferi}, {Moguel}, {Mohapatra}, {Mohite}, {Molina-Ruiz}, {Mondal}, {Mondin}, {Montani}, {Moore}, {Moragues}, {Moraru}, {Morawski}, {More}, {More}, {Moreno}, {Moreno}, {Morisaki}, {Moriwaki}, {Morras}, {Moscatello}, {Mours}, {Mow-Lowry}, {Mozzon}, {Muciaccia}, {Mukherjee}, {Mukherjee}, {Mukherjee}, {Mukherjee}, {Mukund}, {Mullavey}, {Munch}, {Mu{\~n}iz}, {Murray}, {Murray-Dean}, {Muusse}, {Nadji}, {Nagar}, {Nagar}, {Nagarajan}, {Nakamura}, {Nakano}, {Nakano}, {Nakayama}, {Napolano}, {Nardecchia}, {Narikawa}, {Narola}, {Naticchioni}, {Nayak}, {Neil}, {Neilson}, {Nelson}, {Nelson}, {Nery}, {Nesseris}, {Neunzert}, {Ng}, {Ng}, {Nguyen}, {Nguyen}, {Nguyen}, {Nguyen}, {Nguyen Quynh}, {Nichols}, {Nieradka}, {Nishino}, {Nishizawa}, {Nissanke}, {Nitoglia}, {Niu}, {Nocera},
  {Norman}, {North}, {Novak}, {Nu{\~n}o Siles}, {Nurbek}, {Nuttall}, {Oberling}, {O'Dell}, {Oelker}, {Oertel}, {Oganesyan}, {Oh}, {Oh}, {Oh}, {O'Hanlon}, {Ohashi}, {Ohashi}, {Ohkawa}, {Ohme}, {Ohta}, {Oliveira}, {Oliveri}, {Oohara}, {O'Reilly}, {Ormiston}, {Ormsby}, {Orselli}, {O'Shaughnessy}, {O'Shea}, {Oshima}, {Oshino}, {Ossokine}, {Osthelder}, {Ottaway}, {Overmier}, {Pace}, {Pagano}, {Page}, {Pai}, {Pai}, {Pal}, {Palashov}, {P{\'a}lfi}, {Palomba}, {Pan}, {Panda}, {Pang}, {Pannarale}, {Pant}, {Panther}, {Paoletti}, {Paoli}, {Paolone}, {Papalexakis}, {Pappas}, {Parisi}, {Park}, {Parker}, {Pascucci}, {Pasqualetti}, {Passaquieti}, {Passuello}, {Patel}, {Pathak}, {Patra}, {Patricelli}, {Patron}, {Paul}, {Payne}, {Pearce}, {Pedraza}, {Pedurand}, {Pegna}, {Pegoraro}, {Pele}, {Arellano}, {Penn}, {Perego}, {Pereira}, {Perez}, {P{\'e}rigois}, {Perkins}, {Perreca}, {Perri{\`e}s}, {Perry}, {Pesios}, {Petermann}, {Petrillo}, {Pfeiffer}, {Pham}, {Pham}, {Phukon}, {Phurailatpam}, {Piccinni}, {Pichot}, {Piendibene},
  {Piergiovanni}, {Pierini}, {Pierra}, {Pierro}, {Pillant}, {Pillas}, {Pilo}, {Pinard}, {Pineda-Bosque}, {Pinto}, {Piotrzkowski}, {Piotrzkowski}, {Pirello}, {Pitkin}, {Placidi}, {Placidi}, {Planas}, {Plastino}, {Poggiani}, {Polini}, {Pompili}, {Pong}, {Ponrathnam}, {Porcelli}, {Portell}, {Porter}, {Posnansky}, {Poulton}, {Powell}, {Powell}, {Pracchia}, {Pradier}, {Prajapati}, {Prasai}, {Prasanna}, {Pratten}, {Principe}, {Prodi}, {Prokhorov}, {Prosposito}, {Prudenzi}, {Puecher}, {Pullin}, {Punturo}, {Puosi}, {Puppo}, {P{\"u}rrer}, {Qi}, {Quetschke}, {Quinonez}, {Quitzow-James}, {Raab}, {Raaijmakers}, {Radulesco}, {Raffai}, {Rail}, {Raja}, {Rajan}, {Ramirez}, {Ramirez}, {Ramos-Buades}, {Rana}, {Rana}, {Randel}, {Rangnekar}, {Rapagnani}, {Ray}, {Raymond}, {Raza}, {Razzano}, {Read}, {Regimbau}, {Rei}, {Reid}, {Reid}, {Reitze}, {Relton}, {Renzini}, {Rettegno}, {Revenu}, {Reza}, {Rezac}, {Rezaei}, {Ricci}, {Richards}, {Richardson}, {Rijal}, {Riles}, {Riley}, {Rinaldi}, {Robertson}, {Robertson}, {Robinet}, {Rocchi},
  {Rodriguez}, {Rolland}, {Rollins}, {Romanelli}, {Romano}, {Romel}, {Romero}, {Romero-Shaw}, {Romie}, {Ronchini}, {Roocke}, {Rosa}, {Rosauer}, {Rose}, {Rosi{\'n}ska}, {Ross}, {Rossello}, {Roussel}, {Rowan}, {Rowlinson}, {Roy}, {Royzman}, {Rozza}, {Ruggi}, {Ruiz Morales}, {Ruiz-Rocha}, {Ryan}, {Sachdev}, {Sadecki}, {Sadiq}, {Saffarieh}, {Saha}, {Saha}, {Saito}, {Sakai}, {Sakellariadou}, {Sako}, {Sakon}, {Salafia}, {Salces-Carcoba}, {Salconi}, {Saleem}, {Salemi}, {Sall{\'e}}, {Samajdar}, {Sanchez}, {Sanchez}, {Sanchez}, {Sanchis-Gual}, {Sanders}, {Sanuy}, {Saravanan}, {Sarin}, {Sasli}, {Sassi}, {Sassolas}, {Satari}, {Sauter}, {Savage}, {Savant}, {Sawada}, {Sawant}, {Sayah}, {Schaetzl}, {Scheel}, {Scherf}, {Scheuer}, {Schiworski}, {Schmidt}, {Schmidt}, {Schmitz}, {Schnabel}, {Schneewind}, {Schofield}, {Sch{\"o}nbeck}, {Schuler}, {Schulte}, {Schutz}, {Schwartz}, {Scott}, {Scott}, {Seetharamu}, {Seglar-Arroyo}, {Sekiguchi}, {Sellers}, {Sengupta}, {Sentenac}, {Seo}, {Sequino}, {Sergeev}, {Servignat}, {Setyawati},
  {Shaffer}, {Shahriar}, {Shaikh}, {Shams}, {Shao}, {Sharma}, {Chaudhary}, {Shawhan}, {Shcheblanov}, {Sheela}, {Shen}, {Shepard}, {Sheridan}, {Shikano}, {Shikauchi}, {Shimizu}, {Shimode}, {Shinkai}, {Shoemaker}, {Shoemaker}, {Shyamsundar}, {Sider}, {Siegel}, {Sieniawska}, {Sigg}, {Silenzi}, {Singer}, {Singh}, {Singh}, {Singh}, {Singha}, {Sintes}, {Sipala}, {Skliris}, {Slagmolen}, {Slaven-Blair}, {Smetana}, {Smith}, {Smith}, {Smith}, {Soldateschi}, {Somala}, {Somiya}, {Soni}, {Soni}, {Sordini}, {Sorrentino}, {Sorrentino}, {Sotani}, {Soulard}, {Souradeep}, {Sowell}, {Spagnuolo}, {Spencer}, {Spera}, {Spinicelli}, {Srivastava}, {Srivastava}, {Stachie}, {Stachurski}, {Steer}, {Steinlechner}, {Steinlechner}, {Stergioulas}, {Stpierre}, {Strang}, {Stratta}, {Strong}, {Strunk}, {Sturani}, {Stuver}, {Suchenek}, {Sudhagar}, {Sueltmann}, {Sugiyama}, {Suh}, {Sullivan}, {Summerscales}, {Sun}, {Sunil}, {Sur}, {Suresh}, {Sutton}, {Suzuki}, {Suzuki}, {Swinkels}, {Syx}, {Szczepa{\'n}czyk}, {Szewczyk}, {Tacca}, {Tagoshi},
  {Tait}, {Takahashi}, {Takahashi}, {Takamori}, {Takano}, {Takeda}, {Takeda}, {Talbot}, {Talbot}, {Tamaki}, {Tamanini}, {Tanabe}, {Tanaka}, {Tanaka}, {Tanasijczuk}, {Tanioka}, {Tanner}, {Tao}, {Tao}, {Tapia}, {San Mart{\'\i}n}, {Tarafder}, {Taranto}, {Taruya}, {Tasson}, {Teloi}, {Tenorio}, {Terhune}, {Terkowski}, {Themann}, {Thirugnanasambandam}, {Thomas}, {Thomas}, {Thomas}, {Thomas}, {Thompson}, {Thondapu}, {Thorne}, {Thrane}, {Tiwari}, {Tiwari}, {Tiwari}, {Toivonen}, {Tolley}, {Tomaru}, {Tomita}, {Tomura}, {Tonelli}, {Torres-Forn{\'e}}, {Torrie}, {Tosta E Melo}, {Tournefier}, {Trapananti}, {Travasso}, {Traylor}, {Trenado}, {Trevor}, {Tringali}, {Tripathee}, {Troiano}, {Trovato}, {Trozzo}, {Trudeau}, {Tsang}, {Tsang}, {Tse}, {Tso}, {Tsuchida}, {Tsukada}, {Tsutsui}, {Turbang}, {Turconi}, {Turski}, {Tuyenbayev}, {Ubach}, {Ubhi}, {Uchikata}, {Uchiyama}, {Udall}, {Uehara}, {Ueno}, {Unnikrishnan}, {Ushiba}, {Utina}, {Vahlbruch}, {Vaidya}, {Vajente}, {Vajpeyi}, {Valdes}, {Valentini}, {Vallero}, {Valsan}, {van
  Bakel}, {van Beuzekom}, {van Dael}, {van den Brand}, {van den Broeck}, {Vander-Hyde}, {van der Sluys}, {van de Walle}, {van Dongen}, {van Haevermaet}, {van Heijningen}, {Vanosky}, {van Putten}, {van Ranst}, {van Remortel}, {Vardaro}, {Vargas}, {Varma}, {Vas{\'u}th}, {Vecchio}, {Vedovato}, {Veitch}, {Veitch}, {Venneberg}, {Venugopalan}, {Verdier}, {Verkindt}, {Verma}, {Verma}, {Vermeulen}, {Veske}, {Vetrano}, {Vicer{\'e}}, {Vidyant}, {Viets}, {Vijaykumar}, {Villa-Ortega}, {Vina}, {Vincent}, {Vinet}, {Viret}, {Virtuoso}, {Vitale}, {Vocca}, {Voigt}, {von Reis}, {von Wrangel}, {Vorvick}, {Vyatchanin}, {Wade}, {Wade}, {Wagner}, {Walet}, {Walker}, {Wallace}, {Wallace}, {Wang}, {Wang}, {Wang}, {Ward}, {Warner}, {Was}, {Washimi}, {Washington}, {Watada}, {Watarai}, {Watchi}, {Wayt}, {Weaver}, {Weaving}, {Webster}, {Weinert}, {Weinstein}, {Weiss}, {Weller}, {Weller}, {Wellmann}, {Wen}, {We{\ss}els}, {Wette}, {Whelan}, {White}, {Whiting}, {Whittle}, {Wilk}, {Wilken}, {Willetts}, {Williams}, {Williams}, {Williamson},
  {Willis}, {Willke}, {Wipf}, {Woan}, {Woehler}, {Wofford}, {Wong}, {Wong}, {Wong}, {Wright}, {Wu}, {Wu}, {Wu}, {Wysocki}, {Xiao}, {Xu}, {Yadav}, {Yamada}, {Yamamoto}, {Yamamoto}, {Yamamoto}, {Yamamoto}, {Yamamoto}, {Yamashita}, {Yamazaki}, {Yang}, {Yang}, {Yang}, {Yap}, {Yeeles}, {Yelikar}, {Yeung}, {Yokoyama}, {Yokozawa}, {Yoo}, {Yu}, {Yu}, {Yuzurihara}, {Zadro{\.z}ny}, {Zannelli}, {Zanolin}, {Zeeshan}, {Zeidler}, {Zelenova}, {Zendri}, {Zevin}, {Zhang}, {Zhang}, {Zhang}, {Zhang}, {Zhang}, {Zhao}, {Zhao}, {Zhao}, {Zheng}, {Zhong}, {Zhou}, {Zhu}, {Zhu}, {Zimmerman}, {Zucker}, {Zweizig}, {Ligo Scientific Collaboration}, {VIRGO Collaboration}, \& {Kagra Collaboration}}]{2023ApJS..267...29A}
{Abbott}, R., {Abe}, H., {Acernese}, F., {et~al.} 2023, \apjs, 267, 29, \dodoi{10.3847/1538-4365/acdc9f}

\bibitem[{Abbott {et~al.}(2024)}]{LIGOScientific:2023bwz}
Abbott, R., {et~al.} 2024, Astrophys. J., 970, 191, \dodoi{10.3847/1538-4357/ad3e83}

\bibitem[{{Aguado-Barahona} {et~al.}(2022){Aguado-Barahona}, {Rubi{\~n}o-Mart{\'\i}n}, {Ferragamo}, {Barrena}, {Streblyanska}, \& {Tramonte}}]{2022barahona}
{Aguado-Barahona}, A., {Rubi{\~n}o-Mart{\'\i}n}, J.~A., {Ferragamo}, A., {et~al.} 2022, \aap, 659, A126, \dodoi{10.1051/0004-6361/202039980}

\bibitem[{{Birkinshaw}(1989)}]{1989LNP...330...59B}
{Birkinshaw}, M. 1989, in Gravitational Lenses, ed. J.~M. {Moran}, J.~N. {Hewitt}, \& K.-Y. {Lo}, Vol. 330, 59, \dodoi{10.1007/3-540-51061-3_36}

\bibitem[{{Blanchet}(2002)}]{blanchet}
{Blanchet}, L. 2002, Living Reviews in Relativity, 5, 3, \dodoi{10.12942/lrr-2002-3}

\bibitem[{{Blanchet}(2014)}]{Blanchet14}
---. 2014, Living Reviews in Relativity, 17, 2, \dodoi{10.12942/lrr-2014-2}

\bibitem[{{Cahillane} \& {Mansell}(2022)}]{2022ligopl}
{Cahillane}, C., \& {Mansell}, G. 2022, Galaxies, 10, 36, \dodoi{10.3390/galaxies10010036}

\bibitem[{{Capote} {et~al.}(2025){Capote}, {Jia}, {Aritomi}, {Nakano}, {Xu}, {Abbott}, {Abouelfettouh}, {Adhikari}, {Ananyeva}, {Appert}, {Apple}, {Arai}, {Aston}, {Ball}, {Ballmer}, {Barker}, {Barsotti}, {Berger}, {Betzwieser}, {Bhattacharjee}, {Billingsley}, {Biscans}, {Blair}, {Bode}, {Bonilla}, {Bossilkov}, {Branch}, {Brooks}, {Brown}, {Bryant}, {Cahillane}, {Cao}, {Clara}, {Collins}, {Compton}, {Cottingham}, {Coyne}, {Crouch}, {Csizmazia}, {Cumming}, {Dartez}, {Davis}, {Demos}, {Dohmen}, {Driggers}, {Dwyer}, {Effler}, {Ejlli}, {Etzel}, {Evans}, {Feicht}, {Frey}, {Frischhertz}, {Fritschel}, {Frolov}, {Fuentes-Garcia}, {Fulda}, {Fyffe}, {Ganapathy}, {Gateley}, {Gayer}, {Giaime}, {Giardina}, {Glanzer}, {Goetz}, {Goetz}, {Goodwin-Jones}, {Gras}, {Gray}, {Griffith}, {Grote}, {Guidry}, {Gurs}, {Hall}, {Hanks}, {Hanson}, {Heintze}, {Helmling-Cornell}, {Holland}, {Hoyland}, {Huang}, {Inoue}, {James}, {Jamies}, {Jennings}, {Jones}, {Kabagoz}, {Karat}, {Karki}, {Kasprzack}, {Kawabe}, {Kijbunchoo}, {King},
  {Kissel}, {Komori}, {Kontos}, {Kumar}, {Kuns}, {Landry}, {Lantz}, {Laxen}, {Lee}, {Lesovsky}, {Villarreal}, {Lormand}, {Loughlin}, {Macas}, {MacInnis}, {Makarem}, {Mannix}, {Mansell}, {Martin}, {Mason}, {Matichard}, {Mavalvala}, {Maxwell}, {McCarrol}, {McCarthy}, {McClelland}, {McCormick}, {McRae}, {Mera}, {Merilh}, {Meylahn}, {Mittleman}, {Moraru}, {Moreno}, {Mullavey}, {Nelson}, {Neunzert}, {Notte}, {Oberling}, {O'Hanlon}, {Osthelder}, {Ottaway}, {Overmier}, {Parker}, {Patane}, {Pele}, {Pham}, {Pirello}, {Pullin}, {Quetschke}, {Ramirez}, {Ransom}, {Reyes}, {Richardson}, {Robinson}, {Rollins}, {Romel}, {Romie}, {Ross}, {Ryan}, {Sadecki}, {Sanchez}, {Sanchez}, {Sanchez}, {Savage}, {Schaetzl}, {Schiworski}, {Schnabel}, {Schofield}, {Schwartz}, {Sellers}, {Shaffer}, {Short}, {Sigg}, {Slagmolen}, {Soike}, {Soni}, {Srivastava}, {Sun}, {Tanner}, {Thomas}, {Thomas}, {Thorne}, {Todd}, {Torrie}, {Traylor}, {Ubhi}, {Vajente}, {Vanosky}, {Vecchio}, {Veitch}, {Vibhute}, {von Reis}, {Warner}, {Weaver}, {Weiss},
  {Whittle}, {Willke}, {Wipf}, {Wright}, {Yamamoto}, {Zhang}, \& {Zucker}}]{2025ligopl}
{Capote}, E., {Jia}, W., {Aritomi}, N., {et~al.} 2025, \prd, 111, 062002, \dodoi{10.1103/PhysRevD.111.062002}

\bibitem[{\c{C}al\i{}\c{s}kan {et~al.}(2023)\c{C}al\i{}\c{s}kan, Anil~Kumar, Ji, Ezquiaga, Cotesta, Berti, \& Kamionkowski}]{Caliskan:2023zqm}
\c{C}al\i{}\c{s}kan, M., Anil~Kumar, N., Ji, L., {et~al.} 2023, Phys. Rev. D, 108, 123543, \dodoi{10.1103/PhysRevD.108.123543}

\bibitem[{Chen {et~al.}(2024)Chen, Cremonese, Ezquiaga, \& Keitel}]{Chen:2024xal}
Chen, A., Cremonese, P., Ezquiaga, J.~M., \& Keitel, D. 2024, Phys. Rev. D, 110, 123015, \dodoi{10.1103/PhysRevD.110.123015}

\bibitem[{{Chitre} \& {Saslaw}(1989)}]{1989Natur.341...38C}
{Chitre}, S.~M., \& {Saslaw}, W.~C. 1989, \nat, 341, 38, \dodoi{10.1038/341038a0}

\bibitem[{Cremonese {et~al.}(2021)Cremonese, Ezquiaga, \& Salzano}]{Cremonese:2021puh}
Cremonese, P., Ezquiaga, J.~M., \& Salzano, V. 2021, Phys. Rev. D, 104, 023503, \dodoi{10.1103/PhysRevD.104.023503}

\bibitem[{{Dai} {et~al.}(2020){Dai}, {Zackay}, {Venumadhav}, {Roulet}, \& {Zaldarriaga}}]{2020arXiv200712709D}
{Dai}, L., {Zackay}, B., {Venumadhav}, T., {Roulet}, J., \& {Zaldarriaga}, M. 2020, arXiv e-prints, arXiv:2007.12709, \dodoi{10.48550/arXiv.2007.12709}

\bibitem[{{De Filippis} {et~al.}(2004){De Filippis}, {Bautz}, {Sereno}, \& {Garmire}}]{2004ac114}
{De Filippis}, E., {Bautz}, M.~W., {Sereno}, M., \& {Garmire}, G.~P. 2004, \apj, 611, 164, \dodoi{10.1086/422092}

\bibitem[{{D'Orazio} \& {Loeb}(2020)}]{2020PhRvD.101h3031D}
{D'Orazio}, D.~J., \& {Loeb}, A. 2020, \prd, 101, 083031, \dodoi{10.1103/PhysRevD.101.083031}

\bibitem[{{Ezquiaga} {et~al.}(2023){Ezquiaga}, {Hu}, \& {Lo}}]{2023jose}
{Ezquiaga}, J.~M., {Hu}, W., \& {Lo}, R. K.~L. 2023, \prd, 108, 103520, \dodoi{10.1103/PhysRevD.108.103520}

\bibitem[{Ezquiaga \& Zumalac\'arregui(2020)}]{Ezquiaga:2020dao}
Ezquiaga, J.~M., \& Zumalac\'arregui, M. 2020, Phys. Rev. D, 102, 124048, \dodoi{10.1103/PhysRevD.102.124048}

\bibitem[{{Foreman-Mackey} {et~al.}(2013){Foreman-Mackey}, {Hogg}, {Lang}, \& {Goodman}}]{2013emcee}
{Foreman-Mackey}, D., {Hogg}, D.~W., {Lang}, D., \& {Goodman}, J. 2013, \pasp, 125, 306, \dodoi{10.1086/670067}

\bibitem[{{Gond{\'a}n} \& {Kocsis}(2022)}]{2022MNRAS.515.3299G}
{Gond{\'a}n}, L., \& {Kocsis}, B. 2022, \mnras, 515, 3299, \dodoi{10.1093/mnras/stac1985}

\bibitem[{Goyal {et~al.}(2021)Goyal, Haris, Mehta, \& Ajith}]{Goyal:2020bkm}
Goyal, S., Haris, K., Mehta, A.~K., \& Ajith, P. 2021, Phys. Rev. D, 103, 024038, \dodoi{10.1103/PhysRevD.103.024038}

\bibitem[{Goyal {et~al.}(2023)Goyal, Vijaykumar, Ezquiaga, \& Zumalacarregui}]{Goyal:2023uvm}
Goyal, S., Vijaykumar, A., Ezquiaga, J.~M., \& Zumalacarregui, M. 2023, Phys. Rev. D, 108, 024052, \dodoi{10.1103/PhysRevD.108.024052}

\bibitem[{{Gupta} {et~al.}(2024){Gupta}, {Afle}, {Arun}, {Bandopadhyay}, {Baryakhtar}, {Biscoveanu}, {Borhanian}, {Broekgaarden}, {Corsi}, {Dhani}, {Evans}, {Hall}, {Hannuksela}, {Kacanja}, {Kashyap}, {Khadkikar}, {Kuns}, {Li}, {Miller}, {Harvey Nitz}, {Owen}, {Palomba}, {Pearce}, {Phurailatpam}, {Rajbhandari}, {Read}, {Romano}, {Sathyaprakash}, {Shoemaker}, {Singh}, {Vitale}, {Barsotti}, {Berti}, {Cahillane}, {Chen}, {Fritschel}, {Haster}, {Landry}, {Lovelace}, {McClelland}, {J J Slagmolen}, {R Smith}, {Soares-Santos}, {Sun}, {Tanner}, {Yamamoto}, \& {Zucker}}]{2024ligosharp}
{Gupta}, I., {Afle}, C., {Arun}, K.~G., {et~al.} 2024, Classical and Quantum Gravity, 41, 245001, \dodoi{10.1088/1361-6382/ad7b99}

\bibitem[{{Hendriks} {et~al.}(2024{\natexlab{a}}){Hendriks}, {Zwick}, \& {Samsing}}]{2024arXiv240804603H}
{Hendriks}, K., {Zwick}, L., \& {Samsing}, J. 2024{\natexlab{a}}, arXiv e-prints, arXiv:2408.04603, \dodoi{10.48550/arXiv.2408.04603}

\bibitem[{{Hendriks} {et~al.}(2024{\natexlab{b}}){Hendriks}, {Atallah}, {Martinez}, {Zevin}, {Zwick}, {Trani}, {Saini}, {Tak{\'a}tsy}, \& {Samsing}}]{2024arXiv241108572H}
{Hendriks}, K., {Atallah}, D., {Martinez}, M., {et~al.} 2024{\natexlab{b}}, arXiv e-prints, arXiv:2411.08572, \dodoi{10.48550/arXiv.2411.08572}

\bibitem[{Hu \& Wu(2017)}]{10.1093/nsr/nwx116}
Hu, W.-R., \& Wu, Y.-L. 2017, National Science Review, 4, 685, \dodoi{10.1093/nsr/nwx116}

\bibitem[{{Itoh} {et~al.}(2009){Itoh}, {Futamase}, \& {Hattori}}]{2009PhRvD..80d4009I}
{Itoh}, Y., {Futamase}, T., \& {Hattori}, M. 2009, \prd, 80, 044009, \dodoi{10.1103/PhysRevD.80.044009}

\bibitem[{Jana {et~al.}(2023)Jana, Kapadia, Venumadhav, \& Ajith}]{Jana:2022shb}
Jana, S., Kapadia, S.~J., Venumadhav, T., \& Ajith, P. 2023, Phys. Rev. Lett., 130, 261401, \dodoi{10.1103/PhysRevLett.130.261401}

\bibitem[{{Kawamura} {et~al.}(2011){Kawamura}, {Ando}, {Seto}, {Sato}, {Nakamura}, {Tsubono}, {Kanda}, {Tanaka}, {Yokoyama}, {Funaki}, {Numata}, {Ioka}, {Takashima}, {Agatsuma}, {Akutsu}, {Aoyanagi}, {Arai}, {Araya}, {Asada}, {Aso}, {Chen}, {Chiba}, {Ebisuzaki}, {Ejiri}, {Enoki}, {Eriguchi}, {Fujimoto}, {Fujita}, {Fukushima}, {Futamase}, {Harada}, {Hashimoto}, {Hayama}, {Hikida}, {Himemoto}, {Hirabayashi}, {Hiramatsu}, {Hong}, {Horisawa}, {Hosokawa}, {Ichiki}, {Ikegami}, {Inoue}, {Ishidoshiro}, {Ishihara}, {Ishikawa}, {Ishizaki}, {Ito}, {Itoh}, {Izumi}, {Kawano}, {Kawashima}, {Kawazoe}, {Kishimoto}, {Kiuchi}, {Kobayashi}, {Kohri}, {Koizumi}, {Kojima}, {Kokeyama}, {Kokuyama}, {Kotake}, {Kozai}, {Kunimori}, {Kuninaka}, {Kuroda}, {Kuroyanagi}, {Maeda}, {Matsuhara}, {Matsumoto}, {Michimura}, {Miyakawa}, {Miyamoto}, {Miyoki}, {Morimoto}, {Morisawa}, {Moriwaki}, {Mukohyama}, {Musha}, {Nagano}, {Naito}, {Nakamura}, {Nakano}, {Nakao}, {Nakasuka}, {Nakayama}, {Nakazawa}, {Nishida}, {Nishiyama}, {Nishizawa}, {Niwa},
  {Noumi}, {Obuchi}, {Ohashi}, {Ohishi}, {Ohkawa}, {Okada}, {Okada}, {Oohara}, {Sago}, {Saijo}, {Saito}, {Sakagami}, {Sakai}, {Sakata}, {Sasaki}, {Sato}, {Shibata}, {Shinkai}, {Shoda}, {Somiya}, {Sotani}, {Sugiyama}, {Suwa}, {Suzuki}, {Tagoshi}, {Takahashi}, {Takahashi}, {Takahashi}, {Takahashi}, {Takahashi}, {Takahashi}, {Takahashi}, {Akiteru}, {Takano}, {Tanaka}, {Taniguchi}, {Taruya}, {Tashiro}, {Torii}, {Toyoshima}, {Tsujikawa}, {Tsunesada}, {Ueda}, {Ueda}, {Utashima}, {Wakabayashi}, {Yagi}, {Yamakawa}, {Yamamoto}, {Yamazaki}, {Yoo}, {Yoshida}, {Yoshino}, \& {Sun}}]{2011CQGra..28i4011K}
{Kawamura}, S., {Ando}, M., {Seto}, N., {et~al.} 2011, Classical and Quantum Gravity, 28, 094011, \dodoi{10.1088/0264-9381/28/9/094011}

\bibitem[{{Kayser} {et~al.}(1986){Kayser}, {Refsdal}, \& {Stabell}}]{1986A&A...166...36K}
{Kayser}, R., {Refsdal}, S., \& {Stabell}, R. 1986, \aap, 166, 36

\bibitem[{Kocsis {et~al.}(2011)Kocsis, Yunes, \& Loeb}]{Kocsis:2011ka}
Kocsis, B., Yunes, N., \& Loeb, A. 2011, Phys. Rev. D, 84, 024032

\bibitem[{{Liu} {et~al.}(2020){Liu}, {Hu}, {Zhang}, \& {Mei}}]{2020PhRvD.101j3027L}
{Liu}, S., {Hu}, Y.-M., {Zhang}, J.-d., \& {Mei}, J. 2020, \prd, 101, 103027, \dodoi{10.1103/PhysRevD.101.103027}

\bibitem[{{Luo} {et~al.}(2016){Luo}, {Chen}, {Duan}, {Gong}, {Hu}, {Ji}, {Liu}, {Mei}, {Milyukov}, {Sazhin}, {Shao}, {Toth}, {Tu}, {Wang}, {Wang}, {Yeh}, {Zhan}, {Zhang}, {Zharov}, \& {Zhou}}]{2016CQGra..33c5010L}
{Luo}, J., {Chen}, L.-S., {Duan}, H.-Z., {et~al.} 2016, Classical and Quantum Gravity, 33, 035010, \dodoi{10.1088/0264-9381/33/3/035010}

\bibitem[{{Maggiore}(2007)}]{2007maggiore}
{Maggiore}, M. 2007, {Gravitational Waves: Volume 1: Theory and Experiments}, \dodoi{10.1093/acprof:oso/9780198570745.001.0001}

\bibitem[{{More} \& {More}(2022)}]{2022more}
{More}, A., \& {More}, S. 2022, \mnras, 515, 1044, \dodoi{10.1093/mnras/stac1704}

\bibitem[{{Pijnenburg} {et~al.}(2024){Pijnenburg}, {Cusin}, {Pitrou}, \& {Uzan}}]{2024PhRvD.110d4054P}
{Pijnenburg}, M., {Cusin}, G., {Pitrou}, C., \& {Uzan}, J.-P. 2024, \prd, 110, 044054, \dodoi{10.1103/PhysRevD.110.044054}

\bibitem[{{Proust} {et~al.}(2015){Proust}, {Yegorova}, {Saviane}, {Ivanov}, {Bresolin}, {Salzer}, \& {Capelato}}]{2015ac114}
{Proust}, D., {Yegorova}, I., {Saviane}, I., {et~al.} 2015, \mnras, 452, 3304, \dodoi{10.1093/mnras/stv1558}

\bibitem[{{Robertson} {et~al.}(2020){Robertson}, {Smith}, {Massey}, {Eke}, {Jauzac}, {Bianconi}, \& {Ryczanowski}}]{2020robertson}
{Robertson}, A., {Smith}, G.~P., {Massey}, R., {et~al.} 2020, \mnras, 495, 3727, \dodoi{10.1093/mnras/staa1429}

\bibitem[{{Robson} {et~al.}(2019){Robson}, {Cornish}, \& {Liu}}]{2019robson}
{Robson}, T., {Cornish}, N.~J., \& {Liu}, C. 2019, Classical and Quantum Gravity, 36, 105011, \dodoi{10.1088/1361-6382/ab1101}

\bibitem[{{Samsing} {et~al.}(2024{\natexlab{a}}){Samsing}, {Hendriks}, {Zwick}, {D'Orazio}, \& {Liu}}]{2024arXiv240305625S}
{Samsing}, J., {Hendriks}, K., {Zwick}, L., {D'Orazio}, D.~J., \& {Liu}, B. 2024{\natexlab{a}}, arXiv e-prints, arXiv:2403.05625, \dodoi{10.48550/arXiv.2403.05625}

\bibitem[{{Samsing} {et~al.}(2024{\natexlab{b}}){Samsing}, {Zwick}, {Saini}, {D'Orazio}, {Hendriks}, {Mar{\'\i}a Ezquiaga}, {Lo}, {Vujeva}, {Radev}, \& {Yu}}]{2024arXiv241214159S}
{Samsing}, J., {Zwick}, L., {Saini}, P., {et~al.} 2024{\natexlab{b}}, arXiv e-prints, arXiv:2412.14159, \dodoi{10.48550/arXiv.2412.14159}

\bibitem[{{Savastano} {et~al.}(2024){Savastano}, {Vernizzi}, \& {Zumalac{\'a}rregui}}]{2024PhRvD.109b4064S}
{Savastano}, S., {Vernizzi}, F., \& {Zumalac{\'a}rregui}, M. 2024, \prd, 109, 024064, \dodoi{10.1103/PhysRevD.109.024064}

\bibitem[{{Smith} {et~al.}(2019){Smith}, {Bianconi}, {Jauzac}, {Richard}, {Robertson}, {Berry}, {Massey}, {Sharon}, {Farr}, \& {Veitch}}]{2019smith}
{Smith}, G.~P., {Bianconi}, M., {Jauzac}, M., {et~al.} 2019, \mnras, 485, 5180, \dodoi{10.1093/mnras/stz675}

\bibitem[{{Smith} {et~al.}(2023){Smith}, {Robertson}, {Mahler}, {Nicholl}, {Ryczanowski}, {Bianconi}, {Sharon}, {Massey}, {Richard}, \& {Jauzac}}]{smith2023}
{Smith}, G.~P., {Robertson}, A., {Mahler}, G., {et~al.} 2023, \mnras, 520, 702, \dodoi{10.1093/mnras/stad140}

\bibitem[{{Takahashi} \& {Nakamura}(2003)}]{2003ApJ...595.1039T}
{Takahashi}, R., \& {Nakamura}, T. 2003, \apj, 595, 1039, \dodoi{10.1086/377430}

\bibitem[{{Tamanini} {et~al.}(2020){Tamanini}, {Klein}, {Bonvin}, {Barausse}, \& {Caprini}}]{Tamanini2020}
{Tamanini}, N., {Klein}, A., {Bonvin}, C., {Barausse}, E., \& {Caprini}, C. 2020, \prd, 101, 063002, \dodoi{10.1103/PhysRevD.101.063002}

\bibitem[{Tambalo {et~al.}(2023)Tambalo, Zumalac\'arregui, Dai, \& Cheung}]{Tambalo:2022wlm}
Tambalo, G., Zumalac\'arregui, M., Dai, L., \& Cheung, M. H.-Y. 2023, Phys. Rev. D, 108, 103529, \dodoi{10.1103/PhysRevD.108.103529}

\bibitem[{{Tambalo} {et~al.}(2023){Tambalo}, {Zumalac{\'a}rregui}, {Dai}, \& {Cheung}}]{2023PhRvD.108d3527T}
{Tambalo}, G., {Zumalac{\'a}rregui}, M., {Dai}, L., \& {Cheung}, M. H.-Y. 2023, \prd, 108, 043527, \dodoi{10.1103/PhysRevD.108.043527}

\bibitem[{{Tiwari} {et~al.}(2024){Tiwari}, {Vijaykumar}, {Kapadia}, {Fragione}, \& {Chatterjee}}]{tiwari2024}
{Tiwari}, A., {Vijaykumar}, A., {Kapadia}, S.~J., {Fragione}, G., \& {Chatterjee}, S. 2024, \mnras, 527, 8586, \dodoi{10.1093/mnras/stad3749}

\bibitem[{{Vujeva} {et~al.}(2025){Vujeva}, {Mar{\'\i}a Ezquiaga}, {Lo}, \& {Chan}}]{2025arXiv250102096V}
{Vujeva}, L., {Mar{\'\i}a Ezquiaga}, J., {Lo}, R. K.~L., \& {Chan}, J. C.~L. 2025, arXiv e-prints, arXiv:2501.02096, \dodoi{10.48550/arXiv.2501.02096}

\bibitem[{{Wucknitz} \& {Sperhake}(2004)}]{2004PhRvD..69f3001W}
{Wucknitz}, O., \& {Sperhake}, U. 2004, \prd, 69, 063001, \dodoi{10.1103/PhysRevD.69.063001}

\bibitem[{{Xu} {et~al.}(2022){Xu}, {Ezquiaga}, \& {Holz}}]{2022ApJ...929....9X}
{Xu}, F., {Ezquiaga}, J.~M., \& {Holz}, D.~E. 2022, \apj, 929, 9, \dodoi{10.3847/1538-4357/ac58f8}

\bibitem[{{Yang} {et~al.}(2024){Yang}, {Chen}, \& {Cai}}]{2024arXiv241016378Y}
{Yang}, X.-Y., {Chen}, T., \& {Cai}, R.-G. 2024, arXiv e-prints, arXiv:2410.16378, \dodoi{10.48550/arXiv.2410.16378}

\bibitem[{{Zwick} {et~al.}(2023){Zwick}, {Capelo}, \& {Mayer}}]{2023zwick}
{Zwick}, L., {Capelo}, P.~R., \& {Mayer}, L. 2023, \mnras, 521, 4645, \dodoi{10.1093/mnras/stad707}

\end{thebibliography}

\end{document}